\documentclass[nohyper,12pt,letterpaper]{JHEP3}
\usepackage{epsfig}
\def\IZ{{\mathbb Z}}           
\def\IR{{\mathbb R}}

\def\calD{{\cal D}}
\def\calH{{\cal H}}
\def\calN{{\cal N}}
\def\calO{{\cal O}}

\def \tilde{\widetilde}

\def \lr{\leftrightarrow}
\def \eqref#1{(\ref{#1})}

\def \ket#1{\left\vert #1\right\rangle}

\def \exp{\mbox{exp}}

\def \ignoruj#1{}
\def \eqn#1#2{\begin{equation}#2\label{#1}\end{equation}}

\def \gs{g_{\mathrm{string}}}
\def \inte{{\mathrm{interaction}}}

\def \lpl{l_{\mathrm{Planck}}}

\def \da{{\dot a}}
\def \db{{\dot b}}
\def \dc{{\dot c}}
\def \dd{{\dot d}}
\def \CH{\mathfrak {CH}}

\def \vt{\vartheta}
\def \ttheta{{\tilde \theta}}
\def \tSigma{{\tilde \Sigma}}
\def\ignorethis#1{}
\def\be{\begin{equation}}
\def\ee{\end{equation}}
\def\ar#1#2{\begin{array}{#1}#2\end{array}}
\def\bear{\begin{eqnarray}}
\def\eear{\end{eqnarray}}
\def\p{\partial }

\title{Matrix string theory, contact terms,\\
  and superstring field theory}

\author{Robbert Dijkgraaf\\
  Institute for Theoretical Physics \& Korteweg-de Vries Institute for 
Mathematics\\
  University of Amsterdam, Amsterdam, The Netherlands\\
E-mail: \email{rhd@science.uva.nl}}

\author{Lubo\v{s} Motl\\
  Jefferson Physical Laboratory\\
  Harvard University, Cambridge, MA 02138, USA\\
E-mail: \email{motl@feynman.harvard.edu}}

\abstract{In this note, we first explain the equivalence between 
the interaction Hamiltonian of Green-Schwarz light cone gauge
superstring field theory and the twist field formalism known from
matrix string theory. We analyze the role of the large $N$ limit in
matrix string theory, in particular in relation with conformal
perturbation theory around the orbifold SCFT that reproduces
light-cone string perturbation theory. We show how the scaling with
$N$ is directly related to measures on the moduli space of Riemann
surfaces. The scaling dimension $3$ of the Mandelstam vertex as
reproduced by the twist field interaction is in this way related to
the dimension $3(h-1)$ of the moduli space. We analyze the structure
and scaling of the higher order twist fields that represent the
contact terms. We find one relevant twist field at each order. More
generally, the structure of string field theory seems more transparent
in the twist field formalism.  Finally we also investigate the
modifications necessary to describe the pp-wave backgrounds in the
light-cone gauge and we interpret a diagram from the BMN limit as a
stringy diagram involving the contact term.}


\preprint{{\tt hep-th/0309238}\\HEP-UK-0019\\HUTP-03/A063\\ITFA-2003-45}
\begin{document}


\section{Introduction}

Matrix theory provides us with
a fundamental light-cone gauge description of nonperturbative string 
theory in terms of large $N$ matrix models. Although the original BFSS 
matrix model \cite{bfss} covered the 11-dimensional M-theoretical 
background only, it became possible to generalize this formalism 
into other backgrounds as well. See for example the lecture notes
\cite{bilal,banksreview,bigatti,micronotes,taylorreview}.

For instance, the matrix model describing M-theory on $T^k$ for $k\leq
5$ is formulated as the completed $(k+1)$-dimensional maximally
supersymmetric $U(N)$ gauge theory compactified on the dual $T^k$. In
the case $k=4$, the relevant UV completion is the six-dimensional
$(2,0)$ SCFT on $T^5$, and in the case $k=5$ we must deal with the
six-dimensional $(1,1)$ little string theory on $T^5$. These exotic
six-dimensional theories can be described by matrix models
\cite{sixmatrixA,sixmatrixB} or they can be reduced to
four-dimensional theories using the techniques of deconstruction
\cite{lstdecon};
see also \cite{ingodecon}. M-theory on $T^k$ for $k>5$ does not admit a
non-gravitational matrix definition.

The best understood case is $k=1$. In this case, the background of
M-theory on $S^1$ is dual to type IIA string theory. In the limit
where its coupling constant $\gs$ becomes very small, it is possible
to derive the Green-Schwarz light-cone gauge type IIA string field
theory as the appropriate approximation of the matrix model, using the
techniques of matrix string theory \cite{lumodvv,bsdvv,dvv}. Unlike
the light-cone gauge string field theory, the matrix model gives us a
full nonperturbative definition of the stringy dynamics.
It is therefore a fully consistent incarnation of the idea of string bits 
\cite{thornone,thorntwo,thornthree}.

This 1+1-dimensional $U(N)$ gauge theory has the maximal number of 16
supercharges and it contains eight matrix-valued scalar fields $X^i$ that
can be understood as non-Abelian generalizations of the usual eight
transverse coordinates of a string. If one considers the gauge theory with 
the Yang-Mills coupling $g$ (of dimension mass) on a world-sheet cylinder 
of circumference $L$, the type IIA string coupling constant is identified 
with the inverse of the dimensionless gauge coupling
\eqn{grelation}{\gs \sim \frac{1}{gL} \sim\left(R \over 
\lpl\right)^{3/2}.}
The IR limit $L\to\infty$ corresponds to the weak string coupling
limit $\gs\to 0$. Here the supersymmetric Yang-Mills theory becomes
strongly coupled and approaches a superconformal fixed point. The
$U(N)$ gauge symmetry is locally broken down to $U(1)^N$ by the
expectation values of the fields $X^i$. There is a strong evidence
that the IR fixed point is described by the supersymmetrized sigma
model on the symmetric orbifold $\IR^{8N}/S_N$ ({\it i.e.}\ the moduli
space), which can be canonically identified with a free,
second-quantized type IIA string.
In the neighborhood of the fixed
point one hopes to reproduce the standard light-cone perturbative
picture of the interactions by conformal perturbation theory around
this orbifold sigma model.  The perturbation theory in $\gs$ then
corresponds to a strong coupling expansion of the supersymmetric
Yang-Mills theory. In this regime one approximates the matrix string
by an effective Lagrangian density of the form
\eqn{efac}{{\cal L}={\cal L}_{SCFT}+\sum_i \lambda_i \calO_i.}
Here the $\calO_i$ are a set of irrelevant operators in the orbifold model 
that are required to respect space-time supersymmetry and the transverse
rotational group $Spin(8)$ (the R-symmetry), and hopefully---in the large
$N$ limit---the full ten-dimensional super Poincar\'e invariance.
The twist field formalism is more than just a nice set of conventions; it
has been successfully used to calculate
various scattering amplitudes in \cite{frolov,frolovv,frolovvv}.
An operator $\calO_i$ of dimension $d_i$ must be multiplied by a coupling 
constant $\lambda_i$ that scales like $L^{2-d_i}$, which translates into a 
dependence  $\gs^{d_i-2}$ on the string coupling constant. Note that the 
powers $d_i$ are not a priori guaranteed to be integers. However we will 
show that the least irrelevant operators that are invariant under the 
spacetime symmetries have integer total dimensions.

\EPSFIGURE{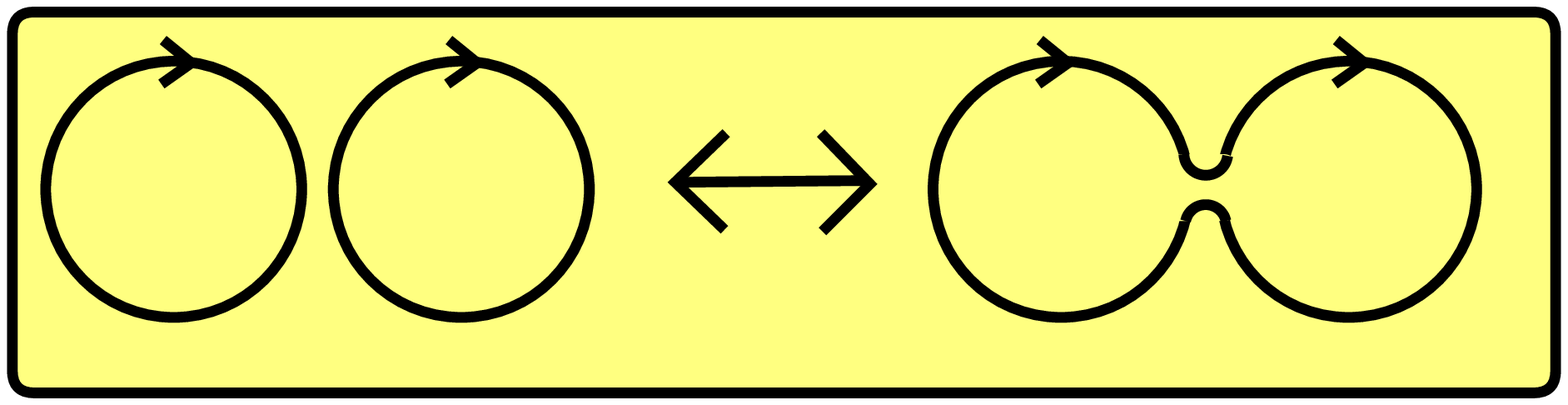,width=84mm}{The $\IZ_2$ twist field induces
the usual splitting/joining interaction between strings, {\it i.e.}\ an
interaction term that is cubic in the string fields.
\label{zetwofig}}

In \cite{dvv} it was shown that the leading irrelevant operator in this 
$\gs$ expansion is given by a specific excited $\IZ_2$ twist field
$\calO_{ij}$ that permutes pairs of two eigenvalue strings $x_i$ and 
$x_j$. In section 2 we show that this DVV twist field exactly reproduces 
the Lorentz-invariant Mandelstam vertex that describes the joining and 
splitting of type II strings in light-cone gauge, including the 
``prefactor''. Since the total scaling dimension of this twist field is 
$3$, this deformation is of first order in the string coupling by the 
scaling argument above.

If we try to go beyond the leading order perturbation, we face a lot of 
issues. Since we flow up in the RG, various more irrelevant terms will 
appear through contact terms, and the question is to which extent the 
super Poincar\'e invariance constrains the effective action \eqref{efac}. 
The hope is that a crucial role is played by the large $N$ limit, and we 
want to analyze this point in more detail in sections \ref{nscaling} and 
\ref{twistfields}.

\subsection{A few more words on matrix string theory}

The generic matrix string configurations locally break the gauge group
$U(N)$ to $U(1)^N$. The coefficient of the commutator terms diverges
for $\gs\to 0$ and therefore dynamics of the gauge theory involves the
moduli space only: in the typical configurations, $X^i(\sigma)$ can be
simultaneously diagonalized for each value of $\sigma$. However the
basis in which they can be diagonalized may undergo a permutation if
$\sigma$ increases by $2\pi$ (the periodicity) because a diagonal
matrix conjugated by a permutation is again a diagonal matrix, and the
symmetric group of permutations is a subgroup of the $U(N)$ gauge
group. Therefore, in the $\gs\to 0$ limit, low energy states of the
gauge theory are divided into sectors classified by a permutation in
$S_N\subseteq U(N)$. In other words, we deal with a two-dimensional
theory on the orbifold
\eqn{orbcft}{\IR^{8N}/S_N}
with the appropriate number of fermionic fields to preserve $\calN=8$
world sheet supersymmetry. The twisted sectors of the orbifold
(\ref{orbcft}) are classified by a permutation. Every permutation may
be factorized into a product of commuting cycles of length $n_i$ and
the corresponding state describes a collection of strings whose
longitudinal momenta equal $p^+_i = n_i / R$. Furthermore, the
orbifold (\ref{orbcft}) requires us to omit the unphysical states
which are not invariant under the $S_N\subseteq U(N)$ gauge
transformations. For a given state, this constraint is nontrivial if
we choose the permutation to be one of the generators of a $\IZ_{n_i}$
group that cyclically permutes a given cycle. In the large $N$ limit,
the ratios $n_i/N$ are kept fixed and $n_i$ are large. The group
$\IZ_{n_i}$ approximates the $U(1)$ group of rigid transformations of
a long string. This implies that only the states that satisfy
$L_0-\tilde L_0=0$ appear in the physical spectrum \cite{dvv}.

Once we consider $\gs$ small but finite, the strings can
interact. Locally a $U(2)_{ij}$ group, originally broken to $U(1)_i
\times U(1)_j$, can get restored.
The detailed physics is a strongly
coupled phenomenon from the gauge theory viewpoint; an instanton 
configuration that might be relevant in this context was suggested
in \cite{nestione,nestitwo}.
But the result of
such a process may include the transposition of the $i$-th and $j$-th
copy of the CFT; $i,j=1\dots N$.  Such a transposition, when added to
the original permutation, can join two cycles into one or split one
cycle into two. This basic mechanism is responsible for stringy
interactions \cite{lumodvv}.

We will often consider the Hamiltonian instead of the Lagrangian; for
the leading order interaction to be described below the identity
${\cal H} = -{\cal L}$ may be applied.  The Hamiltonian for the
orbifold CFT, that approximates the $U(N)$ supersymmetric gauge theory
on $\IR\times S^1$, can be written as
\eqn{hamidvvorig}{P^- = \int_0^1 {\rm d}\sigma
\sum_{m=1}^N
\left[(p_m^i(\sigma))^2
+(x^{\prime i}_m)^2+i\theta'_m\Gamma_9\theta_m(\sigma)
\right]
+P^-_\inte.}  
Here $\theta$ is a 16-component spinor of $SO(9)$,
inherited from the BFSS model, that decomposes into ${\bf 8_s}\oplus
{\bf 8_c}$ under the $SO(8)$ subgroup according to the eigenvalue of
$\Gamma_9$; this eigenvalue decides whether the fermion becomes
left-moving or right-moving. The type IIB D1-brane in the static gauge
correctly reproduces the physics of the type IIA fundamental string in
the light cone gauge.

What about the interactions? The leading term is the least irrelevant operator
preserving the ${\cal N}=(8,8)$ world sheet (or Yang-Mills) supersymmetry:
\eqn{inteeorig}{
{L}_\inte
= \gs
\int_0^L {\rm d}\sigma
\sum_{m<n}^{1\dots N}
(\tau^i\Sigma^i)_{m,n}\otimes (\bar \tau^j\bar\Sigma^j)_{m,n}
\quad + \quad
O(\gs^2)} 
where the excited twist field operator $\tau^i$, $i=1, 2,
\dots 8_v$ and the spin field operators $\Sigma_i$ create a branch cut
in the conformal field theory $CFT_m-CFT_n$, {\it i.e.}\ the sigma
model with coordinates $x_m^i - x_n^i$.  The orthogonal theory
$CFT_m+CFT_n$ (as well as the other $CFT_i$'s for $i\neq m,n$) is not
affected by the permutation of $m$ and $n$. Note that the interaction
factorizes into the left-moving and the right-moving part. The
integrand is an operator of dimension $(3/2,3/2)$. The total dimension
is therefore {\it mass}${}^3$. Because the operator is integrated over
${\rm d}^2\sigma$ to obtain the action, its coefficient must have a
dimension of the world sheet length. Since the only local distance
scale of the gauge theory is $1/g = \gs L$ where $L$ is the
circumference of the $S^1$ in the gauge theory (it is equal to the
inverse mass of the W-bosons that would have to be integrated out in
order to obtain the interaction term), the gauge theory automatically
generates the correct coefficient of (\ref{intee}) proportional to
$\gs$.

Now we want to remind the reader why the operator in (\ref{intee}) has
dimension $(3/2,3/2)$. It is a product of a left-moving and a
right-moving piece and therefore it is sufficient to show that
$\tau^i\Sigma^i$ has dimension $3/2$. Although bosonic string theory
cannot be written as a limit of a consistent gauge theory (because
various supersymmetric cancellations are necessary for the matrix
model to have a spacetime interpretation, {\it i.e.}\ to satisfy the 
cluster
property), it is useful to consider the case of bosonic string theory
first.  In this case, $\tau^i\Sigma^i$ would be simply replaced by
$\sigma$ (and $\bar \tau^i \bar\Sigma^i$ by $\bar \sigma$), the
unexcited twist field:
\eqn{twistun}{\sigma = \prod_{i=1}^{24} \sigma^{(i)}}
It has also dimension $24/16=3/2$ because $1/16$ comes from every
transverse dimension. (The constant $1/16$ equals the difference
between the zero point energy $+1/48$ in the antiperiodic sector and
$-1/24$ in the periodic sector.) In the superstring case we must also
add a spin field because fermions in $CFT_m$ and $CFT_n$ must get
interchanged, too.  But if the fermions $\theta^a$ transform in ${\bf
8_s}$ of $Spin(8)$, their spin fields must transform\footnote{We are
more familiar with the fact that the RNS fermions $\psi^i$
transforming in ${\bf 8_v}$ have spin fields transforming in ${\bf
8_s}\oplus {\bf 8_c}$. These two facts are related by a triality
transformation.}  in ${\bf 8_v}\oplus {\bf 8_c}$, {\it i.e.}\ they are
$\Sigma^i$ and $\Sigma^{\da}$. If we used $\Sigma^\da$, there would be
no chance to contract the spinor index in order to create an $SO(8)$
invariant expression. Therefore we must choose $\Sigma^i$ and its
vector index can be contracted with the vector index of the excited
twist field $\tau^i$, corresponding to the vertex operator of the
state
\eqn{firex}{\alpha^i_{-1/2}\ket{0}_{\mbox{\scriptsize antiperiodic } x^i} 
\lr \tau^i.}
The total dimension of $\tau^i$ is $3/2$: $8/16$ comes from the spin
field $\Sigma_i$, $8/16$ comes from the twist field $\sigma$ and $1/2$
comes from the excitation $\alpha_{-1/2}^i$ in (\ref{firex}).  The
resulting excited twist field $\tau^i\Sigma^i$ may be written as a
supervariation, $\tau^i\Sigma^i=G^\da_{-1/2}(\sigma \Sigma^\da)$.

Furthermore in the case of heterotic strings, we can combine a
left-moving bosonic $\sigma$ with a right-moving supersymmetric
$\bar\tau^i\bar\Sigma^i$. Because both factors have dimension $3/2$,
we again obtain a $(3/2,3/2)$ operator \cite{heterej}.  This fact is
important to preserve the correct scaling \eqref{grelation} of the
interactions also for the heterotic matrix strings
\cite{bamohete,matihorava}.

\subsection{A short review of Green-Schwarz superstring field theory}

Light-cone gauge (super)string field theory is obtained by canonical 
second quantization of the first-quantized quantum mechanics of a single 
string. The amplitudes $\langle u|\psi\rangle$ become operators
$\langle u|\hat\Psi\rangle$ that satisfy the commutation relations
\eqn{secondquan}{[\langle u|\hat\Psi\rangle\, , \,
\langle\hat\Psi| v\rangle]_{\mathrm{grad}} = 
\langle u|v\rangle.}
The annihilation operators $\Psi_u=
\langle u|\hat\Psi\rangle$---and analogously
their Hermitian conjugates, the creation operators $\Psi^\dagger_v=
\langle\hat\Psi| v\rangle$---can be written in terms of various bases
of the first quantized Hilbert space,
for instance a continuous
functional basis, namely as string fields $\hat\Psi[x^i(\sigma),
\theta^a(\sigma)]$ that depend on {\it curves} in a (super)space
much like the fields in point-like particle quantum field theories depend
on {\it points} in a (super)space.

The second-quantized kinematical generators and the free Hamiltonian are 
formally written as the expectation values
in the string field operator-vectors, for example
\eqn{secop}{P^-_{\mathrm{free}}\equiv
H_{\mathrm{free}} = 
\langle \hat\Psi | h_{\mathrm{free}}|\hat \Psi\rangle
=\sum_{u,v}
\langle \hat\Psi |v\rangle\langle v| h_{\mathrm{free}}|u\rangle\langle u|
\hat \Psi\rangle
.}

The interaction Hamiltonian of string field theory annihilates one
string and creates two or vice versa; see figure \ref{zetwofig}. In
the
case of bosonic string field theory, it would have the form
\eqn{bosonich}{\gs\int \calD z_1[\sigma]
\calD z_2[\sigma]
\calD z_3[\sigma]
\Delta[z_3(\sigma)\!-\!z_1(\sigma)\!-\!z_2(\sigma)]
\Psi^\dagger[z_3(\sigma)]
\Psi[z_1(\sigma)]\Psi[z_2(\sigma)]
}
plus the Hermitean conjugate term, where the schematically written
$\Delta$ functional is nonzero only if two parts of the string no. 3
overlap with the string no. 1 or the string no. 2, respectively.  This
continuity condition is automatically satisfied by the twist field
$\sigma$ of the bosonic version of matrix string theory. The bosonic
string interaction vertex contains no prefactors. This reflects the
fact that the interactions of closed bosonic fields can have
non-derivative character (for instance the tachyon potential term
$T^3$).

The vertex in superstring field theory is more complicated. It
contains a ``prefactor'' ${\cal H}$, an operator inserted at the
interaction point.  This prefactor is bilinear in the bosonic fields
$\partial z^i(\sigma_\inte)$.  This reflects the 2-derivative
character of closed superstring field interactions ({\it e.g.}\ $\Phi
R$ and $C \wedge F \wedge F$ in supergravity).
\begin{eqnarray}
&\gs&\int \calD z_1^\oplus[\sigma]
\calD z_2^\oplus[\sigma]
\calD z_3^\oplus[\sigma]\\
&\,&{\cal H}[z^\oplus(\sigma_\inte)]
\Delta[z_3^\oplus(\sigma)\!-\!z_1^\oplus(\sigma)\!-\!z_2^\oplus(\sigma)]
\Psi^\dagger[z_3^\oplus(\sigma)]
\Psi[z_1^\oplus(\sigma)]\Psi[z_2^\oplus(\sigma)]\quad\,
\end{eqnarray}
Here $\calD z^\oplus$ indicates the integral over the whole
superspace, {\it i.e.}\ $\calD^8 z\, \calD^8\Lambda$ where
$\Lambda^a(\sigma)$ will denote eight fermionic coordinates.

The prefactor \cite{brinkgs} is a polynomial
\eqn{asas}{\calH=v^{ij}(\Lambda^a_{\mathrm{reg}})p_{\mathrm{reg}}^i \tilde
p_{\mathrm{reg}}^j.}
Here the function $v^{ij}(\Lambda^a)$ is an octic polynomial in the
fermions near the interaction point, and
\begin{eqnarray}
p_{\mathrm{reg}}^i\sim \sqrt{\epsilon}\partial z^i(\sigma_\inte+\epsilon),
&\qquad&
\tilde p_{\mathrm{reg}}^j\sim\sqrt{\epsilon}\bar\partial
z^i(\sigma_\inte+\epsilon)\\
\Lambda^a_{\mathrm{reg}}\sim \sqrt{\epsilon}\Lambda^a
(\sigma_\inte+\epsilon)
\end{eqnarray}
The operators $\partial z^i(\sigma),\bar\partial
z^i(\sigma),\Lambda^a(\sigma)$ are singular at the interaction point,
and therefore they must be evaluated at $\sigma_\inte+\epsilon$ and
multiplied by $\sqrt{\epsilon}$. The singular behavior can be seen if
we use the coordinate $w$ that is related to $z$ by $z=w^2$:
$\partial_w z^i(w)$ is finite in the $w$-plane (even around $w=0$),
and therefore $\partial_z z^i(z)$ scales as $1/ \sqrt{z}$. A
generalization of the $w$-plane was employed by Wynter \cite{wynter}
to understand the nature of matrix string interactions.

Because an excited version of the twist field $\sigma$ is inserted at
the interaction point, the factor $\sqrt{\epsilon}$ is precisely the
factor needed to compensate $1/\sqrt{z}$ in the OPE of $\sigma$ with
$\partial z^i$. In other words, $p_{\mathrm{reg}}^i$ is proportional
to $\alpha_{-1/2}^i$. Clearly, the factor $p_{\mathrm{reg}}^i$ in
\eqref{asas} transforms $\sigma$ (corresponding to the bare delta
functional) into $\tau^i$, an excited twist field. The factor $\tilde
p_{\mathrm{reg}}^j$ transmutes $\tilde \sigma$ into $\tilde \tau^j$ in a
similar manner.

If we consider the equivalence between the $\sigma$ twist field in
bosonic matrix string theory and the bosonic superstring field
theory's interaction vertex (containing no prefactors) to be a direct
consequence of the topology of this interaction (strings split and
join, as on figure \ref{zetwofig}), the remaining fact to be shown is
that the function $v^{ij}(\Lambda^a_{\mathrm{reg}})$ from \eqref{asas}
describes the correct fermionic twist field
$\Sigma^i\tilde\Sigma^j$. The fermionic zero modes cause the unexcited
fermionic twist field to be degenerate; it has $({\bf 8_v}+{\bf
8_c})\otimes ({\bf 8_v}+{\bf 8_c})$ components: the left-moving part
can be either $\Sigma^i$ or $\Sigma^{\dot a}$ and the right-moving
part can be $\tilde\Sigma^i$ or $\tilde\Sigma^{\dot a}$.  The space of
functions of eight fermions $\Lambda^a$ is also 256-dimensional. Our
task will be to show that $v^{ij}$ corresponds exactly to
$\Sigma^i\tilde\Sigma^j$.


\section{The equivalence of the interaction vertices}

It has become a well-known fact from matrix string theory that the
large $N$ limit of the Hilbert space of the symmetric orbifold CFT
reproduces the Hilbert space of the second-quantized superstring field
theory in the light-cone gauge.  The Hilbert space of large $N$ matrix
string theory contains states with an arbitrary number of strings with
arbitrary values of $P^+$ (while the total $P^+$ is fixed). The states
must be invariant under the orbifold group. This has several
consequences: the unbroken cyclical group $\IZ_k$ commuting with a
cycle of length $k$ becomes $U(1)$ generated by $L_0-\tilde L_0$ when
$k$ is large, and therefore the individual physical strings are
guaranteed\footnote{For finite $k$ this constraint says that
$L_0-\tilde L_0$ must be a multiple of $k$.} to satisfy $L_0=\tilde
L_0$. If several ($M_k$) strings (blocks) carry the same value of $k$,
the unbroken group $\IZ_{M_k}$ exchanging these strings imposes the
(anti)symmetry of the wave function: strings in the same state are
identical, with the correct statistics determined by their spin.

In this section we would like to explicitly show that the $\IZ_2$ twist
field describing the leading perturbation in matrix string theory
coincides with the cubic interaction term in string field theory. The 
reader who is primarily interested in the higher order twist fields should 
skip this section.

Although the easiest theory to derive from the matrix description is
type IIA string theory, we will be focusing on type IIB string
theory. The twist field for type IIB string theory is completely
analogous to that of type IIA string theory; the only difference is
that the chirality of the left-moving fermions in all relevant
formulae is inverted. All $\theta^a$ as well as $\tilde\theta^a$ will
have undotted indices.

This choice allows us to compare the twist field expressions with
the formalism for string field theory by Green, Schwarz, and Brink
\cite{brinkgs} that pairs up eight real left-moving fermions with eight
right-moving fermions into eight complex fermions and their conjugate 
momenta. It has the virtue of keeping the $spin(8)$ symmetry manifest.
The equivalence of the two descriptions of type IIA superstring theory 
then follows from T-duality whose action is simple on both sides.


\subsection{Identification of the unit function of fermions in the DVV 
language}

Let us start with some useful and elementary OPEs of the fermions
$\theta$ and their twist fields $\Sigma$:
\eqn{ope}{\theta^a(z)\Sigma^i(0)\sim \frac{\eta^*}{\sqrt{2z}}
\gamma^{i}_{a\dot a} \Sigma^{\dot a},
\qquad
\theta^a(z)\Sigma^{\dot a}(0)\sim \frac{\eta^*}{\sqrt{2z}}
\gamma^{i}_{a\dot a} \Sigma^{i}.}
We chose an equal phase $\eta^*=\exp(-i\pi/4)$ in both numerators but this phase
must be nontrivial in order to satisfy (\ref{withope}): one can derive this phase
by the requirement that $\theta^{\langle a}\theta^{b\rangle}(z)$ acts as the same
$SO(8)$ generator on $\theta^c(0)$ as well as all the $\Sigma$'s.

We define the antiholomorphic quantities in such a way that they satisfy 
the following OPEs (which differ by $z\leftrightarrow \bar z$
and $\theta\leftrightarrow \tilde \theta$ etc. from (\ref{ope}), without
changing the order of factors):

\eqn{opetwo}{\tilde\theta^a(\bar z)\tilde\Sigma^i(0)\sim 
\frac{\eta^*}{\sqrt{2\bar z}}
\gamma^{i}_{a\dot a} \tilde \Sigma^{\dot a},
\qquad
\tilde\theta^a(\bar z)\tilde\Sigma^{\dot a}(0)\sim \frac{\eta^*}{\sqrt{2z}}
\gamma^{i}_{a\dot a} \tilde\Sigma^{i}.}
We inserted the factors of $1/ \sqrt{2}$ so that the OPEs are consistent 
with
\eqn{withope}{\theta^{a}(z)\theta^{b}(0)\sim \frac{\delta^{ab}}{z},
\quad
\tilde \theta^{a}(\bar z)\tilde\theta^{b}(0)\sim \frac{\delta^{ab}}{\bar 
z}.}

In these calculations, $\theta^a$ is an anticommuting object. The spin 
field $\Sigma^i$ is treated as an anticommuting object, too, because
it corresponds to $\psi^i$ in the RNS formalism. However $\Sigma^{\dot a}$ 
then must be a commuting object because of (\ref{ope}).

The periodic sector of eight pairs of fermions $\theta^a,\tilde\theta^a$ 
contains 
$({\bf 8_v}+{\bf 8_c})\otimes ({\bf 8_v}+{\bf 8_c})$
states (of the supergravity multiplet)
whose vertex operators are
\eqn{verop}{\Sigma^i\tilde\Sigma^{j},\quad
\Sigma^i\tilde\Sigma^{\dot b},\quad
\Sigma^{\dot a}\tilde\Sigma^{j},\quad
\Sigma^{\dot a}\tilde\Sigma^{\dot b}.}
However in the $SO(8)$ invariant formalism for the light-cone gauge 
type IIB
superstring field theory of Green, Schwarz, and Brink \cite{brinkgs}
we must pair the left-moving fermions and the right-moving fermions
into superspace coordinates $\vt^a$ and superspace momenta 
$\lambda^a$:
\eqn{supde}{\vt^a (\sigma) \approx
\eta^*[S_L^a(\sigma)+i S_R^a (\sigma)],\qquad
\lambda^a(\sigma)
\equiv \frac{\partial}{\partial \vt^a(\sigma)}=\eta[S_L^a
(\sigma)-i S_R^a(\sigma)].}
There are eight complex fermionic coordinates $\vt^a$ at each point. 
The fields $S_L,S_R$ are taken to be proportional to
$\theta,\tilde \theta$; also, 
a new symbol $\Lambda$ will be used for 
$\vartheta$ (up to an overall multiplicative factor).
Because the 
three-string vertex contains a prefactor inserted at the interaction 
point, we must study the correspondence between the polynomials of 
$\vt(\sigma_\inte)$ and the operators (\ref{verop}). Which operator 
corresponds to the unit function of $\vt^a(\sigma)$, for example? Because 
the GSB formalism is $SO(8)$ invariant, the unit function must also be
$SO(8)$ invariant. It is not hard to guess that it will contain an equal 
mixture of $\Sigma^i\tSigma^i$ and $\Sigma^{\da}\tSigma^{\da}$:
\eqn{mixture}{1_{\vt}\leftrightarrow
\Sigma^i\tSigma^i - i \Sigma^{\da}\tSigma^{\da}.}
While the overall normalization is somewhat arbitrary (although correlated 
with other conventions), the relative phase $(-i)$ is important to 
guarantee the counterpart of the identity $\partial (\sqrt{2}\eta)/\partial\vt=0$ in 
terms of the OPEs\footnote{We multiplied the equation by two in order to 
get rid of the universal factor $1/2$.}:
\eqn{opesone}{
\begin{array}{rcl}
\eta\sqrt{2}\left(
\theta^a(z)-i\ttheta^a(\bar z)
\right)
\left(
\Sigma^i\tSigma^i(0)-i\Sigma^\db\tSigma^\db(0)
\right)&\sim&z^{-1/2}\gamma^i_{a\da}\Sigma^\da\tSigma^i(0)
+i \bar z^{-1/2}\gamma^i_{a\da}\Sigma^i\tSigma^\da\\
\quad&-&\!i z^{-1/2}\gamma^i_{a\db}\Sigma^i\tSigma^\db(0)  
-\bar z^{-1/2}\gamma^i_{a\db}\Sigma^\db\tSigma^i
\end{array}
}
This vanishes for $z=\bar z$ positive; the positive real $z$-axis is used 
to regulate the quantities that diverge at the interaction point.

\subsection{The remaining polynomials of the fermionic variables}

The remaining polynomials in $\vt^a=\theta^a(z)+i\ttheta^a(\bar z)$ can be 
computed easily. Because (\ref{ope}) and (\ref{opetwo}) imply that 
$\theta(z),\ttheta(\bar z)$ behave in a singular way near the spin field, 
we define new variables
\eqn{Lambda}{\Lambda^a=\sqrt{\frac z2}\theta^a(z)+i\sqrt{\frac{\bar z}2}\ttheta^a(\bar 
z).}
These are clearly related to $\vt^a$, assuming $z$ real and positive.
By a function of $\Lambda^a$, we will mean the limit for $z\to 0$ of the 
OPE of the operator (\ref{Lambda}) with the ``unit'' vertex operator
(\ref{mixture}). One can also check that the derivatives with respect to
$\Lambda^a$ can be represented by
\eqn{derlambda}{\frac{\p}{\p\Lambda^a}=\sqrt{\frac z2}
\theta^a(z)-i\sqrt{\frac{\bar z}2}\ttheta^a(\bar 
z)}
so that the required anticommutators $\{\frac{\p}{\p\Lambda^a},\Lambda^b\}=\delta^{ab}$
are satisfied when acting on the spin fields.

It is natural to start with the linear functions of 
$\theta^a$. In this case, the contributions from (\ref{opesone}) simply 
double:
\eqn{linear}{\Lambda^a \lr \eta^*\gamma^i_{a\da}(\Sigma^\da\tSigma^i
-i\Sigma^i\tSigma^\da).}
Let us act on the previous result with $\Lambda^b$:
\eqn{bilin}{\Lambda^b\Lambda^a\lr
-i\gamma^i_{a\da}\gamma^{j}_{b\da}\Sigma^{\langle j}\tSigma^{i\rangle}
-\gamma^i_{a\da}\gamma^i_{b\db}\Sigma^{\langle 
\db}\tSigma^{\da\rangle}.}
Note that in this $ab$ antisymmetric object
only terms antisymmetric in $ij$ and $\da\db$ appear; they form the 
adjoint 28-dimensional representation of $SO(8)$ in all three cases.
It is straightforward to continue and add $\Lambda^c$:
\eqn{cubic}{\begin{array}{rcl}\Lambda^c\Lambda^b\Lambda^a&\lr&
\frac{-i\eta^*}2\left(
\Sigma^\dc\tSigma^k
+i\Sigma^k\tSigma^\dc
\right)
\left(
\gamma^{\langle k}_{a\da}\gamma^{j\rangle}_{b\da}\gamma^j_{c\dc}
+\gamma^i_{a\langle\da}\gamma^i_{\dc\rangle b}\gamma^k_{c\da}
\right)\\
\quad&=&i\eta^* u^{k\dc}_{abc}\left(
\Sigma^\dc\tSigma^k
+i\Sigma^k\tSigma^\dc
\right)\end{array}
}
The $\gamma$-matrices and the symbols
$u^{k\dc}_{abc}$ and $t^{kl}_{abcd}$ are defined in \cite{brinkgs}.
The quartic polynomial is the last one that we will compute.
\eqn{kvartik}{\begin{array}{rcl}\Lambda^d\Lambda^c\Lambda^b\Lambda^a&\lr&
u^{k\dc}_{\langle abc}
\left(\gamma^{l}_{d\rangle\dc}\Sigma^{(l}\tSigma^{k)}
+i\gamma^k_{d\rangle\dd}\Sigma^\dc\tSigma^\dd
\right)\\
\,&=&t^{kl}_{abcd}\Sigma^k\tSigma^l+i u^{k\dc}_{\langle abc}\gamma^k_{d\rangle\dd}
\Sigma^{\dc}\tSigma^\dd
\end{array}}
The other polynomials are related to those above by the Grassmann
Fourier transform {\it i.e.}\ by adding/removing $\Lambda$'s using the 
epsilon symbol. If we define the operator $\CH$ by
\eqn{ch}{\CH=\left(\Lambda^1-\frac{\p}{\p\Lambda^1}\right)
\left(\Lambda^2-\frac{\p}{\p\Lambda^2}\right)\dots
\left(\Lambda^8-\frac{\p}{\p\Lambda^8}\right)=\prod_{a=1}^8 \sqrt{2\bar z
}\ttheta^a(\bar z),}
it is straightforward to see that $\CH^2=1$ and
\eqn{hodgech}{\begin{array}{rcl}
\CH(\Lambda^{a_8}\dots \Lambda^{a_{9-k}})
&=&\frac{1}{k!}\epsilon^{a_1\dots a_8}\Lambda^{a_1}\dots\Lambda^{a_k}(-1)^k\\
\,&=&
\int
\exp\left(
\sum_{i=1}^8 \Lambda^{i}\Lambda^{\prime i}
\right)
(\Lambda^{\prime a_8}\dots \Lambda^{\prime a_{9-k}})
d\Lambda^{\prime 8} d\Lambda^{\prime 7}\dots d\Lambda^{\prime 1}
\end{array}
}
where $\epsilon^{12345678}=+1$. Note that our formulae follow the equations
in the appendix D of \cite{brinkgs} for $\alpha=2$. More precisely, our $\Lambda$
is related to theirs by
\eqn{vztah}{\Lambda^a = \sqrt{\frac 2\alpha} \Lambda^a_{GSB}.}
How does $\CH$ act on 
operators such as (\ref{verop})? It is an $SO(8)$ invariant operator
with eigenvalues $\pm1$ for $\tSigma^i$ and $\tSigma^\da$, respectively: the equation
(\ref{ch}) implies that $\CH$ acts on the tilded spin fields only and this
operator differs from the chirality operator by a triality transformation.

Finally we consider the polynomials from \cite{brinkgs}:
\eqn{wij}{w^{ij}=\delta^{ij}+\frac{1}{4!}t^{ij}_{abcd}
\Lambda^a\Lambda^b\Lambda^c\Lambda^d+\frac{1}{8!}
\delta^{ij}\epsilon^{abcdefgh}
\Lambda^a\Lambda^b\dots
\Lambda^h}
While $w^{ij}$ is the symmetric part of $v^{ij}=w^{ij}+y^{ij}$,
the quantity $y^{ij}$ is the antisymmetric part:
\eqn{yij}{y^{ij}=-\frac{i}{2}\gamma^{ij}_{ab}\Lambda^a\Lambda^b
-\frac{i}{2\cdot 6!}
\epsilon^{ab\dots h}\Lambda^c\Lambda^d\dots\Lambda^h.}
The effect of the term in (\ref{wij}) proportional to the
$\epsilon$-symbol is to get rid of the $\Sigma^\da\tSigma^\db$
part of (\ref{mixture}) while it doubles
the first part
$\Sigma^j\tSigma^i$. Similarly, the last term in (\ref{yij})
cancels the last term in (\ref{bilin}) but doubles the first term.
The symbol $t^{ij}_{abcd}$ is automatically self-dual:
\eqn{selft}{t^{ij}_{abcd}\equiv \gamma^{ik}_{\langle ab}\gamma^{jk}_{cd\rangle}
=\frac{1}{4!}\epsilon^{abcdefgh}t^{ij}_{efgh}.}
We obtain
\eqn{wiji}{\begin{array}{rcccl}
{w^{ij}}&\lr& (2\delta^{ij}\delta^{kl}+\frac{1}{4!}
t^{ij}_{abcd}t^{kl}_{abcd})
\Sigma^k\tSigma^l&=&+8(\delta^{ik}\delta^{jl}+\delta^{il}\delta^{jk})\Sigma^{k}\tSigma^l
\\
\mbox{and}\quad
y^{ij} &\lr& - \gamma^{ij}_{ab}\gamma^{kl}_{ab}\Sigma^k\tSigma^l
&=&-8(\delta^{ik}\delta^{jl}-\delta^{il}\delta^{jk})\Sigma^{k}\tSigma^l\end{array}}
Therefore the sum admits an easy representation:
\eqn{vijresult}{v^{ij}\equiv w^{ij}+y^{ij} \lr 16\Sigma^j\tSigma^i.}
It is equally straightforward to translate the fermionic functions into the
spin field representation:
\eqn{spinove}{\ar{rcl}
{s_1^{i\da}&=&2\gamma^{i}_{a\da}\Lambda^a + \frac{1}{3\cdot 5!}
u^{i\da}_{abc}\epsilon^{ab\dots h}\Lambda^d\Lambda^e\dots \Lambda^h\\
s_2^{i\da}&=&-\frac 13 u^{i\da}_{abc}\Lambda^a\Lambda^b\Lambda^c
+\frac{2}{7!}\gamma^{i}_{a\da}\epsilon^{ab\dots h}\Lambda^b\Lambda^c
\dots\Lambda^h.}}
Let us consider the combination
\eqn{kombifer}{\frac\eta2(s_1^{i\da}-i s_2^{i\da})\lr
\left(2\gamma^{i}_{a\da}\gamma^{k}_{a\dc}
+\frac{1}3 u^{i\da}_{abc}u_{abc}^{k\dc}\right)\Sigma^{\dc}\tSigma^k
=16\Sigma^{\da}\tSigma^i
}
This operator is important for the interaction part of the dynamical
supersymmetry operator.
In a complete analogy one can also construct the other combination:
\eqn{kombiferr}{-\frac{\eta^*}2(s_1^{i\da}+i s_2^{i\da})\lr
\left(2\gamma^{i}_{a\da}\gamma^{k}_{a\dc}
+\frac{1}3 u^{i\da}_{abc}u_{abc}^{k\dc}\right)\Sigma^{k}\tSigma^\dc
=16\Sigma^{i}\tSigma^\da
}
Type IIB string theory allowed us to use a $Spin(8)$ invariant formalism 
for string field theory; this is also possible for its orientifold, type I 
string theory. Type IIA and heterotic string field theories require us to 
use a formalism that breaks $Spin(8)$. The twist field formalism, on the 
other hand, keeps $Spin(8)$ manifest. The proof of equivalence in the 
other cases could be nevertheless performed in a direct analogy with the 
type IIB proof above.


\section{The large $N$ scaling limit\label{nscaling}}

We now turn to the role of the large $N$ scaling limit  in conformal 
perturbation theory around the $S^N \IR^8$ orbifold model.  Note that in 
contrast with the 't~Hooft limit, where one keeps $g^2N$ fixed and is 
driven to weak coupling $g\to 0$ as $N\to\infty$, in the large $N$ limit 
the dimensionless Yang-Mills coupling constant $gL=1/\gs$ remains fixed. 
Therefore the usual perturbative large $N$ techniques do not apply.

Furthermore, one should take into account  that only Yang-Mills energies 
of the order $1/N$ can give rise to finite spacetime energy in the 
light-cone frame. The truncation to these extremely low-lying states can 
be implemented by a rescaling of the worldsheet time coordinate $\tau$ by 
a factor of $N$:
\eqn{ntau}{\tau\to N\tau.}
The appearance of energies of order $1/N$ in the supersymmetric gauge
theory has an intuitive explanation in the strong coupling IR phase
where the `long string' configurations dominate. In this regime the
matrix-valued coordinates $X^i$ commute almost everywhere, where
they can be simultaneously diagonalized giving $N$ eigenvalue vectors
$x_1^i,\dots , x_N^i$. The long strings are made up of twisted configurations 
of these eigenvalue strings or string bits: the twisted sector 
corresponding to a permutation $p\in S_N\subset U(N)$ describes a
configuration of strings of lengths $n_i$ where
\eqn{cykly}{p=\prod_{i=1}^{N_{\mathrm{cycles}}} (n_i),
\qquad
\sum_{i=1}^{N_{\mathrm{cycles}}} n_i = N}
where $(n_i)$ are the independent cycles of the permutation that
generate $\IZ_{n_i}$ subgroups of $S_N$. These residual groups
$\IZ_{n_i}$ are part of the gauge group that must keep physical states
invariant. This fact imposes the constraints for the individual
strings
\eqn{levelma}{L_0(i) - \tilde L_0 (i) \in n_i \IZ.}
For $N\to\infty$ and $n_i/N$ fixed, $n_i\to\infty$ guarantees that the
finite energy configurations must satisfy $L_0(i)-\tilde L_0(i) = 0$.
If two strings with $n_i=n_j$ are excited in the same state
$|\psi\rangle$, an extra permutation exchanging these two cycles
guarantees that the wave function is (anti)symmetric, according to the
statistics of $|\psi\rangle$.

The worldsheet of the long strings is an $N$-fold cover of the cylinder on 
which the Yang-Mills theory is defined. These covering Riemann surfaces 
have circumference $NL$ and they can support fractional momenta $\sim 
1/N$. The appearance of these long strings is therefore a crucial 
ingredient in our understanding why the large $N$ limit of Matrix theory 
leads to a non-trivial scaling limit of the SYM theory.  Note that, since 
we also scaled the worldsheet time $\tau$ by a factor of $N$, one can 
think of the perturbative string worldsheets as scaled by an overall 
factor of $N$ compared to the SYM cylinder. So scaling in $N$ can be 
thought of in terms of RG flow---a point of view that we will further 
explore.

\subsection{The renormalization group involving $N$\label{nmarginal}}

The leading perturbation is an irrelevant twist field and a natural 
question is why such an irrelevant operator affects physics at very 
long worldsheet distances. The subtlety that makes this twist field 
important is an extra scaling with $N$.

The light-cone Hamiltonian for the orbifold CFT, that approximates
the $U(N)$ supersymmetric gauge theory on $\IR\times S^1$, can be written 
as
\eqn{hamidvv}{P^- = \int_0^L {\rm d}\sigma
\sum_{m=1}^N
\left[(p_m^i(\sigma))^2
+(x^{\prime i}_m)^2+i\theta'_m\Gamma_9\theta_m(\sigma)
\right]
+P^-_\inte.}
Here $\theta$ is a 16-component spinor of $SO(9)$, inherited from the
BFSS model, that decomposes into the eigenvectors of the chirality
matrix $\Gamma_9$, {\it i.e.}\ ${\bf 8_s}\oplus {\bf 8_c}$ under the
$SO(8)$ subgroup. The type IIB D1-brane in the static gauge correctly
reproduces the physics of the type IIA fundamental string in the light
cone gauge.

If the light-like radius $R_-$ becomes infinite and $p^+$ is kept
fixed {\it i.e.}\ $N\to\infty$, the DLCQ treatment becomes ordinary
light-cone gauge quantization and the length of string bits $L$
becomes infinitesimal compared to the total length of the strings
$NL$.  We see that the free term in \eqref{hamidvv} may be rewritten
as an integral over the string(s) of length $NL$:
\eqn{hamidvva}{P^- = \int_0^{NL} {\rm d}\sigma
\left[(p^i(\sigma))^2
+(x^{\prime i})^2+i\theta'\Gamma_9\theta(\sigma)
\right]
+P^-_\inte.}

What about the interactions? The leading term is the least irrelevant
operator preserving the ${\cal N}=(8,8)$ world sheet (or Yang-Mills)
supersymmetry:
\eqn{intee}{P^-_\inte = 
\frac{1}{g}
\int_0^L {\rm d}\sigma
\sum_{m<n}^{1\dots N}
(\tau^i\Sigma^i)_{m,n}\otimes (\bar \tau^j\bar\Sigma^j)_{m,n}
\quad + \quad
O(\gs^2)} 
In this form, the interaction term, resulting from a
strongly coupled dynamics where $U(2)_{m,n}$ (otherwise broken to
$U(1)\times U(1)$) gets restored, there is no $N$-dependence but the
term must be summed over $m,n$. The Yang-Mills coupling constant $g$
(of dimension mass) determines the only local scale of Yang-Mills
theory and the appropriate power is inserted on dimensional grounds
because the twist field has total dimension $3$.  As $N$ becomes
large, the interaction can effectively occur between any two points on
the string(s) and the continuum limit of \eqref{intee} can therefore
be written as a bilocal term
\eqn{inteea}{P^-_\inte = \frac{1}{gL}
\int_0^{NL} {\rm d}\sigma
\int_0^{NL} {\rm d}\sigma'
(\tau^i\Sigma^i)_{(\sigma,\sigma')}\otimes (\bar \tau^j\bar\Sigma^j)_{
(\sigma,\sigma')}\quad + \quad
O(\gs^2)}
Recall that the coefficient of this term is $1/gL=\gs$. The expression 
\eqref{inteea} is locally $N$-independent.

More generally, if the twist field appearing in \eqref{intee} has 
the usual RG 
dimension $d=h+\bar h$, the coefficient will be $1/g^{d-2}$ on dimensional 
grounds. If it connects $w$ indices, {\it  i.e.}\ if the sum contains 
$O(N^w)$ terms (the leading perturbation in \eqref{intee} has $w=2$), then 
the continuum limit, giving a $w$-local (bilocal, trilocal, tetralocal 
etc.) interaction,
requires us add a factor $1/L^{w-1}$. The total 
coefficient replacing $1/gL$ in \eqref{inteea} will be
\eqn{coefa}{\frac{1}{g^{d-2}L^{w-1}}=\frac{\gs^{w-1}}{g^{d-1-w}}}
The power of $\gs$ is thus determined by $w-1$ only: the free Hamiltonian 
has $w=1$ and therefore no $\gs$ dependence. The leading perturbation has 
$w=2$ and is therefore proportional to $\gs$. The $\IZ_k$ twist field, 
permuting $k$ eigenvalue strings, leads to a perturbation of order 
$\gs^{k-1}$. Such a term in the action is generated by the strongly 
coupled gauge theory dynamics in which a group $U(k)$ gets restored.

The denominator in \eqref{coefa} is $g^{d-1-w}$. It is dimensionless
if $d-1-w=0$. In this case, we will say that the twist field is
$N$-marginal and its effects survive the continuum limit. If
$d-1-w>0$, the coefficient $g^{-d+1+w}$ has dimension of a positive
power of length, and therefore this term is $N$-irrelevant in the IR
{\it i.e.}\ for long strings where we can essentially send
$g\to\infty$. $N$-relevant operators would have to have
$d-1-w<0$. Such operators are incompatible with supersymmetry, except
for the (untwisted) mass terms that appear in the light-cone gauge
description of strings in the pp-waves.

In other words, $(d+1-w)=(d-1-w)+2$ is the $N$-corrected dimension
that takes the scaling of $N$ together with worldsheet distances into
account. Only the $N$-marginal operators with $d-1-w=0$ survive the
large $N$ {\it i.e.}\ $g\to\infty$ limit of matrix string theory. In the
section \ref{twistfields} we will see that the leading supersymmetric
$Spin(8)$ invariant twist field from the $\IZ_w$ twisted sector is
$N$-marginal.

Another argument in favor of the condition $d-1-w=0$ in flat space is
the invariance under the boosts generated by $J_{+-}$ that rescale
$P^+$ by $\kappa$ and $P^-$ by $1/\kappa$: the Hamiltonian $P^-$ must
scale like $1/P^+$. The variables $\sigma$ scale just like $P^+$
because the length of the strings represents the light-like momentum.
Therefore $w$ integrals over $\sigma$ in \eqref{inteea} scale like
$(P^+)^w$.  An operator of dimension $d$ scales like $(P^+)^{-d}$. The
product $(P^+)^{-d+w}$ must scale like $(P^+)^{-1}$, and therefore
$d-1-w=0$.

\subsection{Moduli and powers of $N$}

There is actually a direct link between the $N$ scaling of the interaction 
vertex and the moduli of the light-cone diagrams. Consider adding a handle
to a particular diagram by two insertions of the twist field:
\eqn{partin}{\int \calO_{(IJ)}\calO_{(KL)}.}
Each operator $\calO_{(IJ)}$ has dimension $3$ and scales therefore as 
$N^{-3}$. These three factors of $N$ are compensated by two integrals over 
$\sigma$ and one integral over $\tau$.

A product of two such operators \eqref{partin} scales like $N^{-6}$. This 
factor must be cancelled by $6$ explicit factors of $N$. The integral over 
$6$ worldsheet variables has an interpretation in terms of moduli. Adding 
a handle to a Riemann surface generically adds $6$ new real moduli. In the 
light-cone gauge language, they can be interpreted as follows:

\begin{itemize}
\item two time coordinates $\tau,\tau'$ of the interaction vertices
\item one common position of the vertices in $\sigma$; it carries the 
information how $P^+$ is separated between two virtual strings
\item three twist parameters $\Delta \sigma_{1,2,3}$ that implement the 
condition $L_0-\tilde L_0=0$ on two new virtual ``smaller'' strings and 
one new ``bigger'' string
\end{itemize}

We claim that in DLCQ, each of these $6$ real moduli comes with a factor
of $N$, and the total factor of $N^6$ compensates $N^{-6}$ from the
dimensions of the twist fields. The factors of $N$ coming with the
$\tau,\tau',\sigma$ integrals have been explained previously. The
projection to the states satisfying $L_0-\tilde L_0=0$ is approximated by
a $\IZ_{n_i}$ projection and $n_i$ scales like $N$ as well. In fact, in 
this way we could identify the scaling dimension $3$ of the vertex, with 
the number $3$ appearing in the complex dimension $3h-3$ of the moduli 
space of Riemann surfaces of genus $h$.

This analysis can also be extended to include punctures. If the total 
number of external states is $n$, the number of vertices (pairs of pants) 
is given by minus the Euler character $(2h-2+n)$ leading to a total 
scaling dimension of $N^{-(6h-6+3n)}$.  
The number of real moduli is $6h-6+2n$.
There is however an extra factor 
of $N$ for each external state, that implements the $\IZ_{n_i}$ 
level-matching projection. The normalization with $N$ is natural both in 
the orbifold model (viewed as an $S_N$ gauge theory) and on the light-cone 
theory phase space  (where the factors of $N$ are needed to give
$\delta(p^+)$ normalizations of the external states). So, altogether we 
obtain a combinatorial power $N^{6h-6+3n}$ that compensates precisely the 
scaling  of the vertices.

Summarizing: the $N$ scaling weight of correlators in the orbifold SCFT 
agrees with the weight of the corresponding string amplitude viewed as an 
integral over the moduli space of light-cone diagrams.

Let us note that the $\IZ_2$ twist field only has scaling 
dimension $3$ in the case of an 8-dimensional transverse space. This is 
how 
the critical dimension appears in the strong coupling gauge theory. It 
should be compared to the covariant formulation where the coordinate and 
ghost determinants only combine into a proper density over the 
(super)moduli space for the critical space-time dimension.  Note that in 
the subcritical case of  $D<8$ transverse dimensions the string coupling 
constant is $N$-relevant (the dimension is something like $1+D/4$)
and will diverge in the large $N$ limit. This 
might be of relevance for the six-dimensional non-Abelian $(2,0)$ strings, 
that allow a matrix formulation in terms of sigma models on the 
$N$-instanton  moduli space \cite{sixmatrixA}. 

For a transverse $M^4$ (which is either $T^4$ or $K3$) this instanton
moduli space is a hyperK\"a{}hler deformation of a symmetric product,
and part of the above analysis might apply.

Conformal field theory on a deformed symmetric product of many copies
of $M^4$ is the dual description of type IIB string theory on $AdS_3
\times S^3 \times M^4$. The conformal symmetry implies that the
deformation must be marginal in the ordinary sense, not
$N$-marginal. In the pp-wave limit of the symmetric orbifold $S^N
M^4$, one must combine the pp-wave string bit techniques \cite{bmn}
with matrix string theory to understand the full stringy Hamiltonian
\cite{goms}. The reason that the marginal perturbation (the resolution of 
the fixed points of the symmetric orbifold) turns out also to be
$N$-marginal is that a BMN-like mechanism renormalizes the kinetic terms 
in such a way that the worldsheet distances seem to be contracted
by $\gs Q_5$.

\section{Twist fields and contact terms\label{twistfields}}

Let us consider in more detail the large $N$ behavior of operators in
the $\IR^{8N}/S_N$ orbifold. Since the operators naturally factorize
in a product of $\IZ_{n_i}$ factors, it suffices to start with a
single $\IZ_w$ twist field.  Such an operator cyclically permutes the
coordinates $x_1^i, x_2^i, \dots , x_w^i$ of $w$ eigenvalue strings.
In string perturbation theory this operator describes a vertex of
order $w+1$ (or less by an even number) in the string fields.

The twist operator induces twisted boundary conditions for the bosons
\eqn{bosontwi}{x_I^i(\sigma+2\pi)=x^i_{I+1}(\sigma)}
where $I+1$ is computed modulo $w$. The cyclic group element also acts 
on the left-moving as well as the right-moving  fermions $\theta^a_I$
and $\bar \theta^{\dot a}_I$. With periodic (Ramond) boundary conditions, 
the action is
\eqn{ramond}{\theta^a_I(\sigma+2\pi)=\theta_{I+1}^a(\sigma)}
whereas with antiperiodic (Neveu-Schwarz) boundary conditions we have
\eqn{nsw}{\theta^a_I(\sigma+2\pi)=-\theta_{I+1}^a(\sigma).}
Similar relations hold for the right-movers. Note that the Ramond sector 
always has a $16\times 16$ degeneracy, since the linear sum
\eqn{linearsum}{\sum_{I=1}^w  \theta^a_I}
is periodic and gives fermionic zero modes. In the Neveu-Schwarz sector
the fermionic zero modes appear for even $w$ only and have the form
\eqn{linearsumns}{\sum_{I=1}^w (-1)^I \theta^a_I.}
For $w$ odd the Neveu-Schwarz ground state is a singlet. This includes the 
NS vacuum in the untwisted sector $w=1$ as a special case.

It is not difficult to compute the scaling dimensions of the ground states 
in such $\IZ_w$ twisted sectors. First of all, both for the bosons and the 
fermions the $\IZ_w$ action can be diagonalized by taking complex linear 
combinations. (In the case of fermions, this only leaves the subgroup
$SU(4)\times U(1)$ of $spin(8)$ manifest.)

The bosonic twist field that implements a twist with eigenvalue
$e^{2\pi i k/w}$ where $k=0,1,\dots w-1$, is well-known to have the
total (left+right) scaling dimension $Dk(w-k)/2w^2$ with $D$ the
dimension of the transversal space (in our case of type II string
theory, $D=8$). For the fermionic twist field with the same twist
$e^{2\pi i k/w}$ ($k$ is an integer or half-integer for the NS or R
sector respectively) we find dimension $Dm^2/2w^2$ where
$m=\min(k,w-k)$.

Summing up all possible eigenvalues for $D=8$ we obtain a scaling 
dimension $w$ for the R ground states and
\eqn{addingup}{
d=\left\{\begin{array}{ll}
w,&w \mbox{\,\,\,even}\\
w-1/w,&w \mbox{\,\,\,odd}
\end{array}\quad
\right\rangle\quad\mbox{for the NS ground states.}
}
We are however only interested in those states that are  invariant under 
the supersymmetry and $Spin(8)$. Although the NS ground states for $w$ odd 
are $Spin(8)$ invariant, they are not supersymmetric.

The least irrelevant  invariant twist field $\calO_{(w)}$ can be 
constructed analogously to the $\IZ_2$ operator in \cite{dvv}. 
Because of supersymmetry at the leading order in $\gs$,
it must be 
written as a supersymmetric descendent of a NS state with the
correct indices  $\dot a, b$
\eqn{desce}{\calO_{(w)}=G^{\dot a}_{-1/2}
\bar G_{-1/2}^b \calO_{(w)}^{\dot a b}.}
For $w=2$ we obtain the leading perturbation (DVV twist field). More 
generally for $w$ even, $\calO_{(w)}^{\dot a b}$ is exactly one of the 
degenerate ground state twist fields. However for $w$ odd \eqref{desce} 
must be completed by a definition of $\calO_{(w)}^{\dot a b}$ because the
ground state $\calO_{(w)}$ is non-degenerate in this case. The right 
prescription for $\calO_{(w)}^{\dot a b}$ can be easily guessed in analogy 
with the case $w=1$ where
\eqn{nisone}{\partial x^i\bar\partial x_i
= G_{-1/2}^{\dot a} \bar G_{-1/2}^{b} \gamma^i_{a\dot a}
\gamma_{i,b \dot b}  \theta^a \bar\theta^{\dot b}.}
Because this is again a descendant, it is supersymmetric  up to a total 
derivative. For general odd $w$ we can therefore define the least 
irrelevant operator appearing in \eqref{desce} as
\eqn{defii}{ \calO_{(w)}^{\dot a b} = 
\gamma^i_{a\dot a}\gamma_{i,b \dot b}
\theta_{-1/2w}^a\bar\theta^{\dot b}_{-1/2w}
|0\rangle}
where $|0\rangle$ denotes the unique $\IZ_w$ NS ground state. 
One might also consider another operator of the same dimension as
\eqref{defii}, namely $\theta_{-1/2w}^b\bar\theta^{\dot a}_{-1/2w}
|0\rangle$, but it would transform incorrectly under supersymmetry;
the operator $\gamma^i_{a\dot a}\gamma_{i,b \dot b}$ that projects
the spinors onto their self-dual part seems to play a crucial role.
The invariant operator $\calO_{(w)}$ that is reproduced in this way
has again total scaling dimension $w+1$, and therefore is $N$-marginal
as described in the subsection \ref{nmarginal}. The operators then
lead to the $\gs^{w-1}$ dependence on the string coupling
constant. The $S_N$ invariance implies that the term must be summed
over the whole conjugacy class:
\eqn{conjclass}{\gs^{w-1} \int \mathrm{d}\sigma
\sum_{I_1<\dots I_w} \calO_{(I_1\dots I_w)}.}
All other supersymmetric and $Spin(8)$ invariant operators in the
twisted sector are $N$-irrelevant, and therefore will become
unimportant in the large $N$ limit. Intuitively it is not too
surprising because such operators are obtained by acting with
$\partial x^i$ on the operator $\calO$ that leads to an extra factor
of $1/N$ because of the long strings.

Note that the unique operator \eqref{desce} replaces a whole family of 
different operators that one would have to insert in string field theory.
For example the $\IZ_3$ vertex induces the interactions from figure
\ref{zethreefig} and the higher-order diagrams would lead to an even 
larger set of interactions.

We mention that the operators of dynamical spacetime supercharges
$Q^{-\dot a}$ and $\bar Q^{-b}$ contain terms similar to
\eqref{desce}, but without one of the worldsheet supersymmetric
excitation:
\eqn{desceq}{
\calO_{(w)}^{\dot a}=
\bar G_{-1/2}^b \calO_{(w)}^{\dot a b},
\quad
\bar\calO_{(w)}^{b}=G^{\dot a}_{-1/2}
\calO_{(w)}^{\dot a b}.}
It would be interesting to compute the actual numerical coefficients
of all the operators $\calO_{(w)}$ in $P^-,Q^{-\dot a},\bar
Q^{-b}$. We believe that the expansion is non-polynomial, much like
other closed string theory actions as well as their low energy limit,
{\it i.e.}\ the Einstein-Hilbert action. The coefficients of the
contact terms must be usually taken to be infinite,
$1/\epsilon^\alpha$, and the circumference of the Yang-Mills cylinder
$L$ will play the role of the natural worldsheet cutoff $\epsilon$ in
our case.

\subsection{An example: $\IZ_2$ and $\IZ_3$ twist fields and pp-waves}

The $\IZ_2$ twist field $\calO_{(IJ)}$ has the interpretation of the 
three-string joining/splitting vertex (see the figure
\ref{zetwofig}). The higher order 
twist fields describe {\it spacetime} contact terms that are known to 
appear in light-cone gauge perturbation theory 
\cite{greensiteone,greensitetwo,greenseiberg,greensitethree}.

\EPSFIGURE{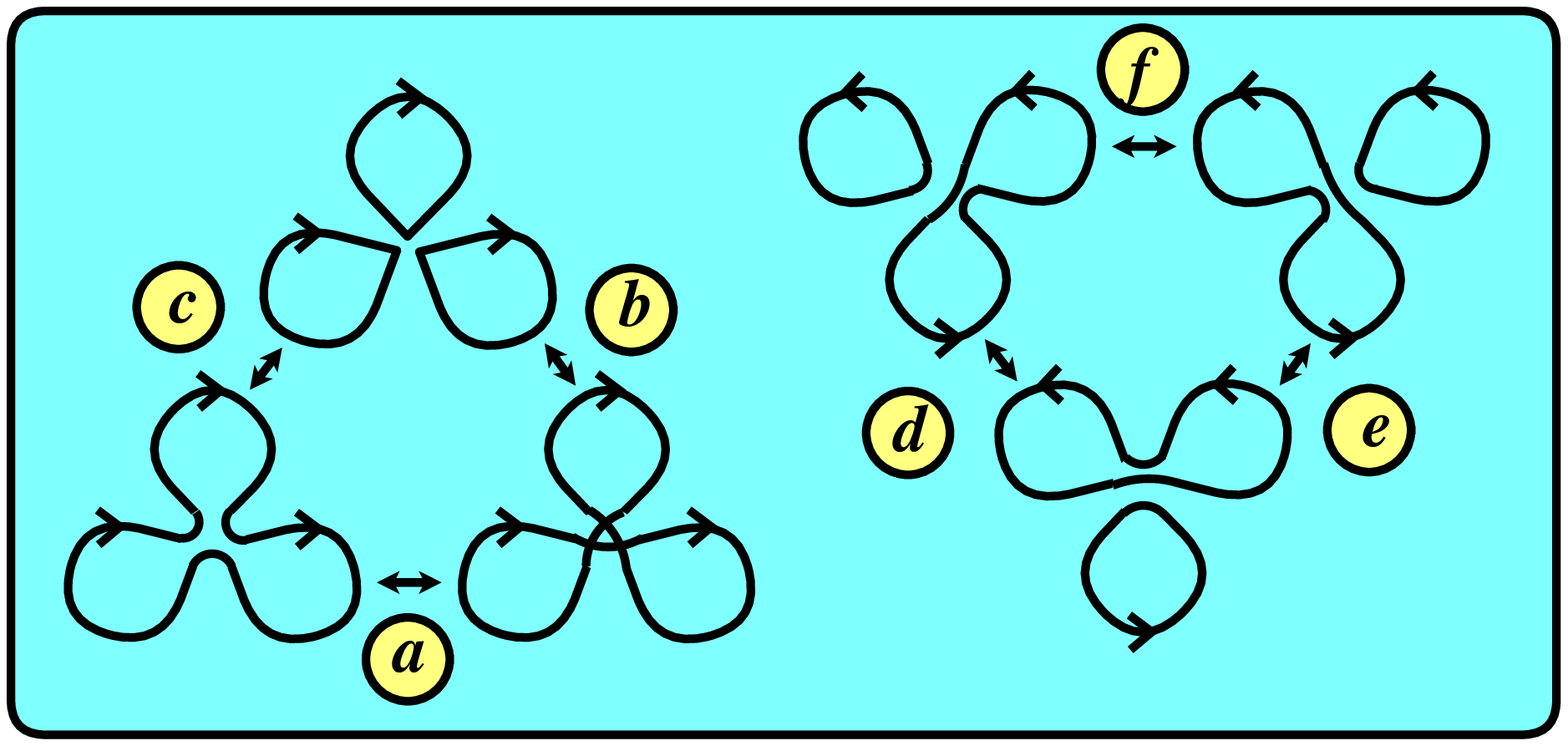,width=144mm}{The $\IZ_3$ twist field induces
quadratic {\bf (a)} and quartic {\bf (b,c,d,e,f)} (depending on the choice 
of the twisted sector) contact
interactions of the string fields. The local form of the trilocal
interaction is identical in all cases. Higher order twist fields give rise 
to ever larger number of different string field interaction terms.
\label{zethreefig}}

For example, locally the same $\IZ_3$ vertex (that is an effective 
description of physical phenomena resulting from a $U(3)$ symmetry 
restoration) generates a quartic as
well as a quadratic contact interaction (see figure \ref{zethreefig})
in the string fields that scales like $\gs^2$. The quadratic
interaction is necessary \cite{greensiteone,greensitetwo,greensitethree}
to cancel the $\gs^2$ self-energy of the
supergraviton ground state in the second-order of the old-fashioned
quantum perturbation theory which is known to be negative for the
ground state:
\eqn{perthe}{E^{(2)}_n=\sum_{m}^{m\neq n}
\frac{|V_{mn}|^2}{E_n^{(0)}-E_m^{(0)}}.}

\EPSFIGURE{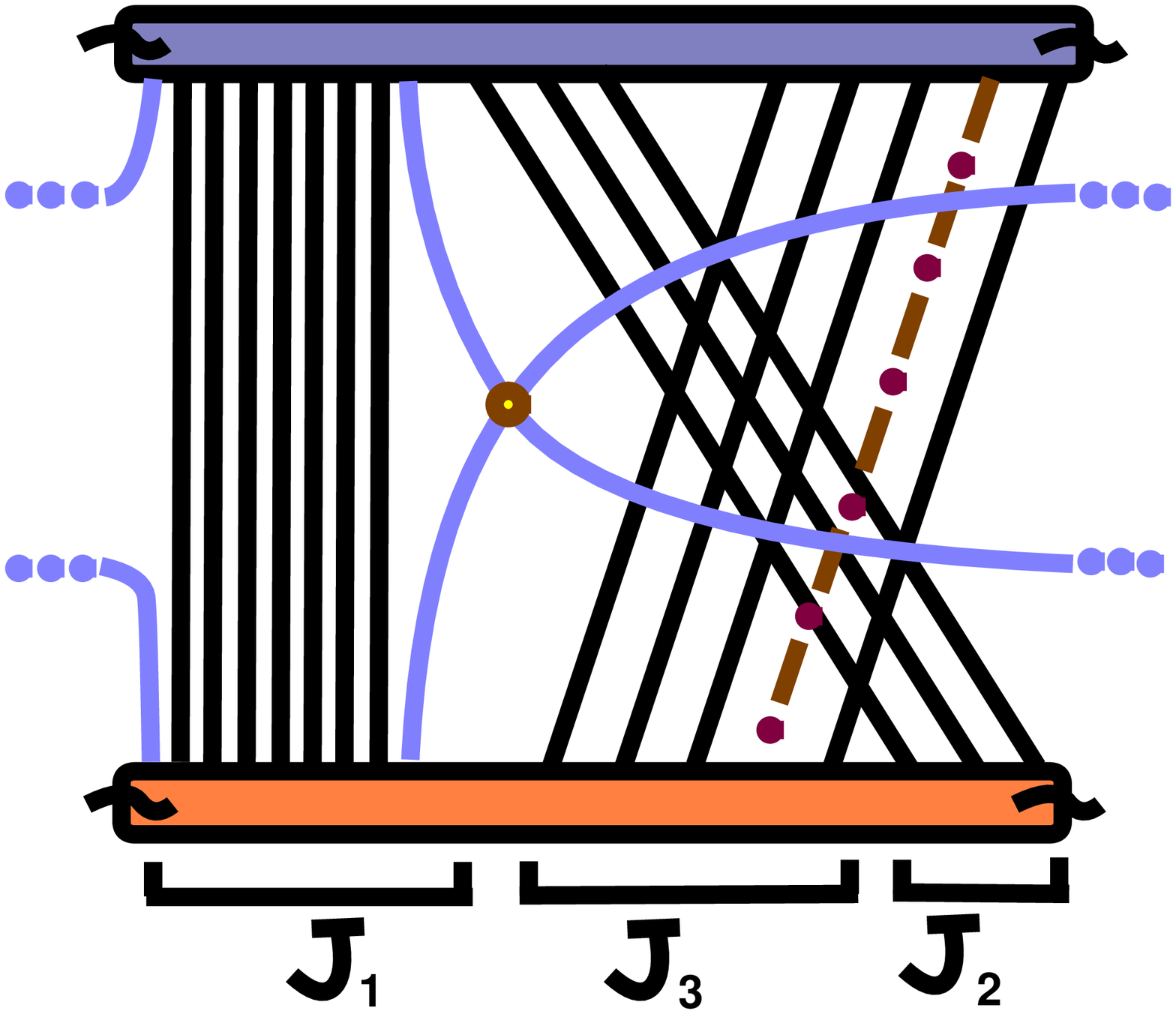,width=84mm}{The non-nearest neighbor diagram
from \cite{constable} contributing to a self-energy of a string has the 
form of the contact interaction from figure {\bf \ref{zethreefig}(a)}.
\label{nonnearr}}

This is just a special case of a more general requirement that the super 
Poincar\'{e} algebra is closed at higher orders in $\gs$; this requirement 
forces us to add higher order terms into the Hamiltonian and the dynamical 
supersymmetry generators.

A simple argument showing that these $\gs^2$ terms are inevitable is
the following: the ground state energy ({\it i.e.}\ the mass of the graviton
multiplet) always acquires a {\it negative} shift in the second order
perturbation theory, and an explicit positive shift proportional to
$\gs^2$ (see figure \ref{zethreefig}(a)) is needed to compensate this
second-order correction and keep the graviton massless.

We can also see that the non-nearest neighbor diagrams of
\cite{constable} (see also \cite{staudacher}) depicted on figure
\ref{nonnearr} that contribute to the self-energy of the states in the
pp-wave background have exactly the structure of the quadratic $\IZ_3$
contact interaction. (A string is divided to three pieces that get
rearranged.)  Note that the pp-wave deformation of string theory is
relevant, and therefore does not affect the UV physics on the
worldsheet. However, the exact analysis of the self-energy is more
complicated because of the operator mixing effects
\cite{mixingboston,mixingreich,mixingqm,mixingdilat}.
A general proposal to identify all the Feynman diagrams that are 
responsible for the contact terms appeared in section 5 of 
\cite{umutcontact}.

\subsection{Composite operators}

Up to now we only analyzed the operators in a single $\IZ_n$ twisted 
sector. The full orbifold also contains twisted superselection sectors 
that are products of such cyclic permutations. In that case one simply 
takes the products of the operators of the individual factors. However, 
these composite operators are always $N$-irrelevant.

Consider a simple example: a conjugacy class of type $(IJ)(KL)$ with 
$I,J,K,L$ all different, consisting of two elementary transpositions. The 
obvious guess for the least-irrelevant operator in this sector is
\eqn{obvious}{\calO_{(IJ)(KL)}
=\calO_{(IJ)}\calO_{(KL)}.}
Note that the OPE of the factors on the right hand side of
\eqref{obvious} is non-singular because the twist fields act on
different coordinates. It has total scaling dimension $6$ and
combinatorial weight $4$, and therefore scales as $N^{-1}$. It is
therefore $N$-irrelevant. One could also try other combinations, such
as $\calO^{\dot a b}_{(IJ)} \calO_{(KL)}^{\dot a b}$ of weight
four. This is $spin(8)$ invariant, but not supersymmetric.

In summary: it is only the irreducible $\IZ_n$ vertices that survive the 
large $N$ scaling.

\subsection{Contact terms on the worldsheet}

The $\IZ_2$ twist field $\calO_{(IJ)}$ has an interpretation as the 
three-string joining/splitting vertex. The higher order twist fields 
describe {\it spacetime} contact terms, that are known to appear in 
light-cone  perturbation theory. They are necessary for the supersymmetry 
algebra to be closed at higher orders in $\gs$.

In fact, these terms also appear as {\it worldsheet} contact terms in the 
conformal perturbation theory. For example, in the OPE of two $\IZ_2$ 
twist fields the $\IZ_3$ twist field can appear through
\eqn{zthreeappear}{\calO_{(IJ)}(z)
\calO_{(JK)}(w)\sim\frac{1}{|z-w|^2}\calO_{(IJK)}(w),
\qquad I\neq K.}
The singularity is logarithmic (just as in the OPE of marginal operators), 
so that in the RG flow the $\IZ_3$ contact term $\calO_{(IJK)}(w)$ is 
reproduced with a finite coefficient. So the effective action contains a 
term
\eqn{efftri}{\gs^2 \int 
\sum_{I<J<K} \calO_{(IJK)}.}
One easily verifies that in all OPEs between these least irrelevant 
invariant operators the singularities  are of the above type. (Roughly 
because they are descendents of chiral primary fields.) Since all such
operators are $N$-marginal, this contact term algebra is preserved in the 
large $N$ limit. 

Note that in this way the $\IZ_n$ twist fields for $n>2$ are made out of 
contact terms between the fundamental $\IZ_2$ vertices, these higher order 
interactions also respect the 10-dimensional Lorentz symmetry!

\section{Conclusions and outlook}

We believe that the twist field formulation of perturbative string theory 
in the light-cone gauge is more natural and more fundamental
than the usual language of second-quantized string field theory. A single 
and simple twist field at each order in $\gs$ gives rise to many 
polynomial interactions of the string fields. 
Yet, it is still straightforward to translate the expressions involving
twist fields into those involving string fields.
Instead of complicated 
Neumann coefficients, one can study simple OPEs in a conformal field 
theory. A modified version of the renormalization group includes the 
$N$-scaling as well and helps us to understand the large $N$ limit that is 
responsible for the light-like decompactification limit of Matrix theory.

It could be interesting to:
\begin{itemize}
\item determine the precise coefficients of the $\IZ_{n}$ twist fields 
from the requirement that the super Poincar\'{e} algebra is closed; are 
all the coefficients non-zero? A recent investigation of the pp-waves
\cite{nastaherman} seems to indicate that all terms beyond $\gs^2$ vanish;
\item compute the matrix elements of the contact terms between general
string states;
\item study the divergence of the string coupling expansion of the 
light-cone Hamiltonian itself; does it diverge in the same sense as the 
action of closed {\it covariant} string field theories?
\item calculate some explicit loop diagrams and understand how the 
singular coefficients of the contact terms arise from our finite $\IZ_{n}$ 
twist fields;
\item try to derive the explicit $\IZ_n$ twist field perturbations from 
the $U(n)$ Yang-Mills theory more directly, perhaps by some sort of 
instanton calculation;
\item check that the dimensions of $N$-marginal operators are integers 
also in the case of heterotic string theories;
\item extend the twist field formalism to open strings; the 
splitting/joining interaction vertex for the open strings should mimic the 
structure of the left-moving part of the closed string vertex.
\end{itemize}

\acknowledgments

We would like to express our special gratitude to
Steve Shenker for his collaboration at the early
stages of this work.
We are also grateful to 
Tom Banks,
Rami Entin,
Simeon Hellerman,
Gautam Mandal,
Shiraz Minwalla,
Andrew Neitzke, 
Marcus Spradlin,
Herman Verlinde,
and
Anastasia Volovich
for useful discussions.
The research of R.D. 
is partly supported by FOM and CMPA grant of the University of Amsterdam.
The work of L.M. was supported in part by Harvard DOE grant
DE-FG01-91ER40654 and the Harvard Society of Fellows.



\begin{thebibliography}{19}        

\bibitem{bfss}
T.~Banks, W.~Fischler, S.~H.~Shenker and L.~Susskind,
{\it ``M theory as a matrix model: A conjecture,''}
Phys.\ Rev.\ D {\bf 55}, 5112 (1997)
[arXiv:hep-th/9610043].

\bibitem{bilal}
A.~Bilal,
{\it ``M(atrix) theory: A pedagogical introduction,''}
Fortsch.\ Phys.\  {\bf 47}, 5 (1999)
[arXiv:hep-th/9710136].

\bibitem{banksreview}
T.~Banks,
{\it ``Matrix theory,''}
Nucl.\ Phys.\ Proc.\ Suppl.\  {\bf 67}, 180 (1998)
[arXiv:hep-th/9710231].

\bibitem{bigatti}
D.~Bigatti and L.~Susskind,
{\it ``Review of matrix theory,''}
arXiv:hep-th/9712072.

\bibitem{micronotes}
R.~Dijkgraaf, E.~Verlinde and H.~Verlinde,
{\it ``Notes on matrix and micro strings,''}
Nucl.\ Phys.\ Proc.\ Suppl.\  {\bf 62}, 348 (1998)
[arXiv:hep-th/9709107].

\bibitem{taylorreview}
W.~Taylor,
{\it ``M(atrix) theory: Matrix quantum mechanics as a fundamental 
theory,''}
Rev.\ Mod.\ Phys.\  {\bf 73}, 419 (2001)
[arXiv:hep-th/0101126].

\bibitem{sixmatrixA}
O.~Aharony, M.~Berkooz, S.~Kachru, N.~Seiberg and E.~Silverstein,
{\it ``Matrix description of interacting theories in six dimensions,''}
Adv.\ Theor.\ Math.\ Phys.\  {\bf 1}, 148 (1998)
[arXiv:hep-th/9707079].

\bibitem{sixmatrixB}
O.~Aharony, M.~Berkooz, S.~Kachru and E.~Silverstein,
{\it ``Matrix description of (1,0) theories in six dimensions,''}
Phys.\ Lett.\ B {\bf 420}, 55 (1998)
[arXiv:hep-th/9709118].

\bibitem{lstdecon}
N.~Arkani-Hamed, A.~G.~Cohen, D.~B.~Kaplan, A.~Karch and L.~Motl,
{\it ``Deconstructing (2,0) and little string theories,''}
arXiv:hep-th/0110146.

\bibitem{ingodecon}
I.~Kirsch and D.~Oprisa,
{\it ``Towards the deconstruction of M-theory,''}
arXiv:hep-th/0307180.

\bibitem{lumodvv}
L.~Motl,
{\it ``Proposals on nonperturbative superstring interactions,''}
arXiv:hep-th/9701025.

\bibitem{bsdvv}
T.~Banks and N.~Seiberg,
{\it ``Strings from matrices,''}
Nucl.\ Phys.\ B {\bf 497}, 41 (1997)
[arXiv:hep-th/9702187].

\bibitem{dvv}
R.~Dijkgraaf, E.~Verlinde and H.~Verlinde,
{\it ``Matrix string theory,''}
Nucl.\ Phys.\ B {\bf 500}, 43 (1997)
[arXiv:hep-th/9703030].

\bibitem{thornone}
R.~Giles and C.~B.~Thorn,
{\it ``A Lattice Approach To String Theory,''}
Phys.\ Rev.\ D {\bf 16}, 366 (1977).

\bibitem{thorntwo}
C.~B.~Thorn,
{\it ``On The Derivation Of Dual Models From Field Theory,''}
Phys.\ Lett.\ B {\bf 70}, 85 (1977).

\bibitem{thornthree}
C.~B.~Thorn,
{\it ``On The Derivation Of Dual Models From Field Theory 2,''}
Phys.\ Rev.\ D {\bf 17}, 1073 (1978).

\bibitem{frolov}
G.~E.~Arutyunov and S.~A.~Frolov,
{\it ``Virasoro amplitude from the S(N) R**24 orbifold sigma model,''}
Theor.\ Math.\ Phys.\  {\bf 114}, 43 (1998)
[arXiv:hep-th/9708129].

\bibitem{frolovv}
G.~E.~Arutyunov and S.~A.~Frolov,
{\it ``Four graviton scattering amplitude from S(N) R**8 supersymmetric  
orbifold sigma model,''}
Nucl.\ Phys.\ B {\bf 524}, 159 (1998)
[arXiv:hep-th/9712061].

\bibitem{frolovvv}
G.~Arutyunov, S.~Frolov and A.~Polishchuk,
{\it ``On Lorentz invariance and supersymmetry of four particle scattering  
amplitudes in S(N) R**8 orbifold sigma model,''}
Phys.\ Rev.\ D {\bf 60}, 066003 (1999)
[arXiv:hep-th/9812119].

\bibitem{nestione}
G.~Bonelli, L.~Bonora and F.~Nesti,
{\it ``Matrix string theory, 2D SYM instantons and affine Toda systems,''}
Phys.\ Lett.\ B {\bf 435}, 303 (1998)
[arXiv:hep-th/9805071].

\bibitem{nestitwo}
G.~Bonelli, L.~Bonora, F.~Nesti and A.~Tomasiello,
{\it ``Matrix string theory and its moduli space,''}
Nucl.\ Phys.\ B {\bf 554}, 103 (1999)
[arXiv:hep-th/9901093].

\bibitem{wynter}
T.~Wynter,
{\it ``Gauge fields and interactions in matrix string theory,''}
Phys.\ Lett.\ B {\bf 415}, 349 (1997)
[arXiv:hep-th/9709029].

\bibitem{heterej}
S.~J.~Rey,
{\it ``Heterotic M(atrix) strings and their interactions,''}
Nucl.\ Phys.\ B {\bf 502}, 170 (1997)
[arXiv:hep-th/9704158].

\bibitem{bamohete}
T.~Banks and L.~Motl,
{\it ``Heterotic strings from matrices,''}
JHEP {\bf 9712}, 004 (1997)
[arXiv:hep-th/9703218].

\bibitem{matihorava}
P.~Ho\v{r}ava,
{\it ``Matrix theory and heterotic strings on tori,''}
Nucl.\ Phys.\ B {\bf 505}, 84 (1997)
[arXiv:hep-th/9705055].

\bibitem{brinkgs}
M.~B.~Green, J.~H.~Schwarz and L.~Brink,
{\it ``Superfield Theory Of Type II Superstrings,''}
Nucl.\ Phys.\ B {\bf 219}, 437 (1983).

\bibitem{bmn}
D.~Berenstein, J.~Maldacena and H.~Nastase,
{\it ``Strings in flat space and pp waves from $N=4$ super Yang Mills,''}
arXiv:hep-th/0202021.

\bibitem{goms}
J.~Gomis, L.~Motl and A.~Strominger,
{\it ``PP-wave / CFT(2) duality,''}
JHEP {\bf 0211}, 016 (2002)
[arXiv:hep-th/0206166].

\bibitem{greensiteone}
J.~Greensite and F.~R.~Klinkhamer,
{\it ``New Interactions For Superstrings,''}
Nucl.\ Phys.\ B {\bf 281}, 269 (1987).

\bibitem{greensitetwo}
J.~Greensite and F.~R.~Klinkhamer,
{\it ``Contact Interactions In Closed Superstring Field Theory,''}
Nucl.\ Phys.\ B {\bf 291}, 557 (1987).

\bibitem{greenseiberg}
M.~B.~Green and N.~Seiberg,
{\it ``Contact Interactions In Superstring Theory,''}
Nucl.\ Phys.\ B {\bf 299}, 559 (1988).

\bibitem{greensitethree}
J.~Greensite and F.~R.~Klinkhamer,
{\it ``Superstring Amplitudes And Contact Interactions,''}
Nucl.\ Phys.\ B {\bf 304}, 108 (1988).

\bibitem{staudacher}
C.~Kristjansen, J.~Plefka, G.~W.~Semenoff and M.~Staudacher,
{\it ``A new double-scaling limit of N = 4 super Yang-Mills theory and 
PP-wave strings,''}
Nucl.\ Phys.\ B {\bf 643}, 3 (2002)
[arXiv:hep-th/0205033].

\bibitem{constable}
N.~R.~Constable, D.~Z.~Freedman, M.~Headrick, S.~Minwalla, L.~Motl,
A.~Postnikov and W.~Skiba,
{\it ``PP-wave string interactions from perturbative Yang-Mills theory,''}
JHEP {\bf 0207}, 017 (2002)
[arXiv:hep-th/0205089].

\bibitem{mixingboston}
N.~R.~Constable, D.~Z.~Freedman, M.~Headrick and S.~Minwalla,
{\it ``Operator mixing and the BMN correspondence,''}
JHEP {\bf 0210}, 068 (2002)
[arXiv:hep-th/0209002].

\bibitem{mixingreich}
N.~Beisert, C.~Kristjansen, J.~Plefka, G.~W.~Semenoff and M.~Staudacher,
{\it ``BMN correlators and operator mixing in N = 4 super Yang-Mills 
theory,''}
Nucl.\ Phys.\ B {\bf 650}, 125 (2003)
[arXiv:hep-th/0208178].

\bibitem{mixingqm}
N.~Beisert, C.~Kristjansen, J.~Plefka and M.~Staudacher,
{\it ``BMN gauge theory as a quantum mechanical system,''}
Phys.\ Lett.\ B {\bf 558}, 229 (2003)
[arXiv:hep-th/0212269].

\bibitem{mixingdilat}
N.~Beisert, C.~Kristjansen and M.~Staudacher,
{\it ``The dilatation operator of N = 4 super Yang-Mills theory,''}
Nucl.\ Phys.\ B {\bf 664}, 131 (2003)
[arXiv:hep-th/0303060].

\bibitem{umutcontact}
U.~Gursoy,
{\it ``Predictions for pp-wave string amplitudes from perturbative SYM,''}
arXiv:hep-th/0212118.

\bibitem{gsw}
M.B.~Green, J.H.~Schwarz, E.~Witten,
{\it ``Superstring theory,''} 2 volumes,
Cambridge University Press 1987

\bibitem{joebig}
J.~Polchinski,
{\it ``String theory,''}
2 volumes, Cambridge University Press 1998


\bibitem{maldacena}
J.~Maldacena,
{\it ``The large $N$ limit of superconformal field theories and
supergravity,''}
Adv.\ Theor.\ Math.\ Phys.\  {\bf 2}, 231 (1998)
[Int.\ J.\ Theor.\ Phys.\  {\bf 38}, 1113 (1998)]
[arXiv:hep-th/9711200].

\bibitem{ssethi}
S.~Sethi and L.~Susskind,
{\it ``Rotational invariance in the M(atrix) formulation of type IIB theory,''}
Phys.\ Lett.\ B {\bf 400}, 265 (1997)
[arXiv:hep-th/9702101].

\bibitem{nastaherman}
J.~Pearson, M.~Spradlin, D.~Vaman, H.~Verlinde and A.~Volovich,
{\it ``Tracing the string: BMN correspondence at finite J**2/N,''}
JHEP {\bf 0305}, 022 (2003)
[arXiv:hep-th/0210102].

\end{thebibliography}
\end{document}